\documentclass[12pt]{iopart}
\usepackage{hyperref}
\expandafter\let\csname equation*\endcsname\relax

\expandafter\let\csname endequation*\endcsname\relax

\usepackage{amsmath}
\usepackage{amssymb}
\usepackage{multirow}
\usepackage{graphicx}
\usepackage{bm}
\usepackage{comment}

\begin{document}

\title[Spherically Symmetric Potentials in Quadratic $f(R)$ Gravity]{Spherically Symmetric Potentials in Quadratic $f(R)$ Gravity}

\author{Roger Anderson Hurtado}
\address{Observatorio Astronómico Nacional, Universidad Nacional de Colombia,\\
Av. Carrera 30 45-03 Edif. 476, Bogotá, Colombia}
\ead{rahurtadom@unal.edu.co}
\vspace{10pt}
\begin{indented}
\item[\today]
\end{indented}

\begin{abstract}
We study the gravitational potential generated by static, spherically symmetric matter distributions in a quadratic $f(R)$ gravity model. In the weak-field regime, the linearized field equations lead to a fourth-order modified Poisson equation whose solutions contain Newtonian and Yukawa-type contributions. Imposing regularity at the origin and asymptotic flatness uniquely fixes the integration constants, yielding potentials fully determined by the mass density. Analytical expressions are derived for several classical profiles, including Plummer, Hernquist, and Navarro–Frenk–White (NFW), as well as for new analytic density models introduced in this work. The dependence on the quadratic gravity parameter $\alpha$ is analyzed, and the Newtonian limit of General Relativity is consistently recovered as $\alpha \to \infty$. As an application, circular velocity curves are computed and compared with the observed rotation curve of NGC 3198. A chi-squared analysis shows that the linearized quadratic $f(R)$ model provides improved fits relative to the Newtonian case in the inner and intermediate galactic regions $r \lesssim 30$ kpc, while predicting a decline at larger radii due to Yukawa suppression.
\end{abstract}

%
\section{\label{sec:level1}Introduction}
The quest to understand the fundamental nature of gravity has led to the development of numerous alternative theories, among which $f(R)$ gravity stands out for its mathematical elegance and phenomenological richness \cite{Sotiriou_2006,Sotiriou:2008rp,De_Felice_2010,Clifton_2012,Shankaranarayanan_2022,Hurtado_2023,hurtadoRo}. In these models, the gravitational action is generalized to depend on an arbitrary function of the Ricci scalar, $f(R)$, allowing for modifications to Einstein's General Relativity (GR) that can address outstanding problems in cosmology and astrophysics \cite{Capozziello:2004km,Ivanov_2022,Vilenkin,Nojiri}, such as the nature of dark energy \cite{Nojiri,Amendola_2007b,Caprini_2016,oikonomou,ODINTSOV2023137988,Chatterjee_2024} and the dynamics of galactic rotation curves \cite{zhang,martins,stabile,naik}. Quadratic models represent a minimal extension with significant theoretical and observational consequences \cite{Starobinsky:1980te,joachim,Zhdanov_2024}. While the cosmological implications of quadratic $f(R)$ gravity have been extensively studied \cite{Giamalas,Asaka_2016,Gegenberg_2018,Chakraborty_2021,Nashed_2023}, much less attention has been paid to its impact on local gravitational potentials generated by realistic matter distributions.
\\
In particular, static spherically symmetric sources—relevant for galaxies, dark matter halos, and compact objects—serve as ideal testbeds to explore the structure of the modified potential and its sensitivity to matter distributions. These systems provide a controlled environment where the interplay between gravity and the underlying mass profile can be studied in detail, offering insights into both fundamental theory and astrophysical phenomena. Standard mass profiles such as Plummer \cite{Plummer}, Hernquist \cite{hernquist,Baes_2002}, and Navarro–Frenk–White (NFW) \cite{NFW,Navarro_1997} have been widely used in astrophysics due to their empirical success in modeling luminous and dark matter components \cite{Jing_2000,Merritt_2006}, capturing the observed dynamics of stellar systems and the rotation curves of galaxies with remarkable accuracy \cite{Oman_2015,kuzio,Block}. Each of these profiles exhibits distinct inner and outer density behaviors, which in turn influence the resulting gravitational potential and its modifications under alternative gravity theories.
\\
Quadratic extensions of General Relativity have been extensively studied as consistent higher-derivative theories of gravity. In particular, Lü, Perkins, Pope, and Stelle \cite{Lu} analyzed the most general parity, even quadratic action, including both $R^2$ and Weyl-squared terms, and derived static, spherically symmetric solutions in both the exact and linearized regimes, establishing the theoretical consistency of the theory and examining several source configurations in the weak-field limit. However, a systematic analytical treatment of realistic astrophysical density profiles (e.g., the Hernquist or NFW profiles) within such modified gravity frameworks remains largely unexplored. Addressing this gap is essential, because gravitational modifications introduce new features, such as Yukawa-type corrections and altered asymptotic behavior, that can leave observable imprints on galactic dynamics, gravitational lensing, and the structure of compact objects. A comprehensive analytical understanding not only clarifies the theoretical implications of these corrections but also provides indispensable tools for interpreting astrophysical data and for testing gravity beyond the standard Newtonian regime.
\\
In contrast, the present work focuses on the pure $f(R)=R+\alpha R^2$ sector and provides a systematic analysis of the Newtonian limit gravitational potential generated by a broad class of spherical mass distributions, including astrophysically motivated profiles and new analytic models, with direct application to galactic rotation curves.

This paper is organized as follows. Section \ref{sec:level2} focuses on deriving the exact solutions of the linearized fourth-order field equations in a quadratic $f(R)$ gravity model, under the assumption of static spherical symmetry. Section \ref{sec:level3} provides an analytical solution to this equation, with integration constants determined by imposing appropriate physical conditions on the gravitational potential. In Section \ref{sec:level4}, the modified potential is computed and analyzed for various spherical mass distributions, including both classical profiles and newly proposed models. Section \ref{sec:level5} then computes the corresponding rotation curves for selected density profiles and confronts them with the measured rotation curve of NGC 3198, providing a direct astrophysical test of the model. Finally, Section \ref{sec:level6} discusses the main conclusions of this work.

\section{\label{sec:level2}Field equations in $f(R)$ theory}
We begin by considering $f(R)$ gravity as a generalization of the Einstein-Hilbert action in which the Ricci scalar $R$ is replaced by a function $f(R)$. This action takes the form
\begin{equation}
    S=\frac{1}{2\kappa}\int f(R)\sqrt{-g}d^4x+S_M,
\end{equation}
where $\kappa=8\pi$ in geometrized units, $g$ is the determinant of the metric tensor $g_{\mu\nu}$, and $S_M$ denotes the matter and energy contribution. Varying this action with respect to the metric, yields the modified field equations \cite{Sotiriou:2008rp,De_Felice_2010}
\begin{equation}\label{fieldequations}
    FR_{\mu\nu}-\frac{1}{2}fg_{\mu\nu}-\nabla_{\mu}\nabla_{\nu}F+g_{\mu\nu} \Box F=\kappa T_{\mu\nu}, 
\end{equation}
where $f=f(R)$, $F=df/dR$, $\nabla_\mu$ denotes the covariant derivative, $\Box=\nabla_\sigma\nabla^\sigma=g^{\sigma\rho}\nabla_\sigma\nabla_\rho$ is the d'Alembert operator, and $T_{\mu\nu}$ is the stress-energy tensor, defined by
\begin{equation}
    T_{\mu\nu}=\frac{-2}{\sqrt{-g}}\frac{\delta S_M}{\delta g^{\mu\nu}}.
\end{equation}
The terms $\nabla_\mu\nabla_\nu F$ and $\Box F$ in Eq. (\ref{fieldequations}), introduce fourth-order derivatives of the metric, distinguishing $f(R)$ gravity from second-order GR. Tracing this equation reveals a dynamical scalar degree of freedom, $F$, encoding the modified gravitational interaction.
In this work, we consider the specific functional form
\begin{equation}\label{model}
    f(R)=-\frac{4\pi}{\beta}\left(4\gamma+2\alpha^2R+R^2\right),
\end{equation}
where $\alpha$, $\beta$ and $\gamma$ are constant parameters, which encompasses several important limits of gravitational theories. In the limit $\alpha^2\to\infty$, $\beta/\alpha^2=-8\pi$ and $\gamma/\beta=\Lambda/(8\pi)$, the action reduces to GR with a cosmological constant $\Lambda$. Moreover, setting $\gamma=0$ recovers pure GR. Interestingly, for $\alpha^2=3 m^2$, $\beta/\alpha^2=-8\pi$, $\gamma=0$, the model corresponds to the well-known Starobinsky quadratic gravity model \cite{Starobinsky:1980te},
\begin{equation}\label{starobinsky}
    f(R)=R+\frac{1}{6m^2}R^2,
\end{equation}
where the parameter $m$ sets the mass scale of the scalaron, the additional scalar degree of freedom arising from the $R^2$ term, which is associated with the energy scale of inflation at the early universe, and acts as a coupling constant for higher-order curvature effects. In particular, the measured amplitude of primordial curvature perturbations in the Cosmic Microwave Background (CMB) fixes $m\approx10^{-5}M_{pl}$, (where $M_{pl}$ is the reduced Planck mass), consistent with Planck data \cite{planckcollaboration1,planckcollaboration2}. Now, using Eq. (\ref{model}), field equations can be written as
\begin{equation}\label{fieldequa}
    \alpha^2 G_{\mu\nu}-\left(\gamma+\frac{1}{4}R^2\right)g_{\mu\nu}+\left(g_{\mu\nu}\Box-\nabla_{\mu}\nabla_{\nu}+R_{\mu\nu}\right)R=-\beta T_{\mu\nu}.
\end{equation}
where we have used the Einstein tensor
\begin{equation}
    G_{\mu\nu}=R_{\mu\nu}-\frac{1}{2}Rg_{\mu\nu}.
\end{equation}
To study weak-field gravity in this framework, we linearize Eq. (\ref{fieldequa}) around a Minkowski background metric $\eta_{\mu\nu}=\text{diag}(-1,1,1,1)$, by writing 
\begin{equation}\label{approximation}
    g_{\mu\nu}=\eta_{\mu\nu}+h_{\mu\nu},\qquad h_{\mu\nu}\ll 1,
\end{equation}
where $h_{\mu\nu}$ represents a small perturbation (dimensionless in geometrized units), and work in the Newtonian limit where gravitational fields are static, therefore time derivatives can be neglected ($\partial_th_{00}=0$) and the d'Alembert operator $\Box$ reduces to the spatial Laplacian $\nabla^2$, moreover, in this regime we consider matter sources such that $T_{00}=\rho\gg |T_{ij}|$. Focusing on the time-time component of the perturbation and by using the trace reverse tensor 
\begin{equation}
    \bar h_{\mu\nu}=h_{\mu\nu}-\frac{1}{2}\eta_{\mu\nu}h,
\end{equation}
the linearized equation takes the form
\begin{equation}
    (\alpha^2 \nabla^2+2 \gamma)\bar{h}_{00}+(\nabla^4+\gamma)\bar{h}=2\left(\beta\rho+\gamma\right),
\end{equation}
by neglecting the linear term due to Eq. (\ref{approximation}) and reverting to the original perturbation $h_{\mu\nu}$, we arrive at the modified Poisson equation
\begin{equation}\label{poisson}
    \alpha^2\nabla^2h_{00}-\nabla^{4}h_{00}=\beta \rho+\gamma,
\end{equation}
this is a fourth-order elliptic differential equation for the gravitational potential $h_{00}$ in the Newtonian limit. The term $\alpha^2\nabla^2h_{00}$ is associated to the standard Newtonian behavior, while the biharmonic operator $\nabla^{4}$ reflects the higher-derivative nature of the underlying $f(R)$ theory and introduces modifications relevant at short distances or high curvature. The source includes both, energy density, $\rho$, and a constant term, $\gamma$, acting as an effective background source related to the cosmological constant in the weak-field regime.
\\
The structure of the linearized field equations and the emergence of Yukawa-type corrections are consistent with previous analyses of quadratic gravity in the weak-field regime \cite{Lu}, where the massive scalar and spin-2 modes were identified.


\section{\label{sec:level3}Solution for Spherically Symmetric Sources}
We now proceed to solve the field equation (\ref{poisson}) in the presence of a static, spherically symmetric source. In this case, the metric perturbation depends only on the radial coordinate $r$, and we neglect the cosmological constant by setting $\gamma=0$ to isolate local gravitational effects. Therefore, Eq. (\ref{poisson}) reduces to
\begin{equation}\label{HT}
    \alpha^2\left(2h_{00}'(r)+r h_{00}''(r)\right)-4h_{00}^{(3)}(r)-r h_{00}^{(4)}(r)=\beta r\rho(r),
\end{equation}
this fourth-order ODE admits an analytical solution through successive integration, for this we can use the following identity
\begin{equation}\label{hH}
    \frac{d}{dr}\left(r^2\frac{dh_{00}(r)}{dr}\right)=r\frac{d^2}{dr^2}\left[r h_{00}(r)\right],
\end{equation}
which, when applied recursively, transforms the equation into
\begin{equation}\label{eqgreen1}
    \frac{d^2}{dr^2}\left(\alpha^2 r h_{00}(r)-\frac{d^2}{dr^2}\left[r h_{00}(r)\right]\right)=\beta r \rho(r),
\end{equation}
this form is particularly convenient for integration, allowing the use of definite integrals with carefully chosen limits. The resulting solution is
\begin{equation}\label{HTintegral}
    h_{00}(r)=-\frac{\beta e^{\alpha r}}{r}\int_r^{\infty}e^{-2\alpha u}\int_0^ue^{\alpha q}\int_0^q\int_p^{\infty}s\rho(s)\, ds\,  dp\, dq\, du -\frac{h_{00}(0)}{2\alpha}\frac{e^{-\alpha r}}{r},
\end{equation}
which, for physically viable solutions, must satisfy two key boundary conditions: asymptotic flatness, requiring that the potential decays at spatial infinity, \(\lim_{r\to\infty} r h_{00}(r) = 0\); and regularity at the origin, \(h_{00}(0) = \text{finite}\). These conditions directly constrain the form of the solution. The requirement of asymptotic flatness eliminates exponentially growing modes, ensuring that \(h_{00}(r)\) falls off as \(\mathcal{O}(1/r)\) or faster. Consequently, all higher radial derivatives exhibit the expected decay: \(r h_{00}'(r) \sim \mathcal{O}(1/r)\), \(h_{00}''(r) \sim \mathcal{O}(1/r^3)\), and so on.  
\\
Near the origin, we assume the perturbation is regular and can be expanded as a Taylor series:  
\[
h_{00}(r) \simeq h_{00}(0) + r h_{00}'(0) + \frac{1}{2} r^2 h_{00}''(0) + \mathcal{O}(r^3).
\]  
Imposing symmetry at the origin implies \(h_{00}'(0) = 0\). Together with finiteness of \(h_{00}(0)\), this determines the remaining freedom in the solution.
\\
Taking into account that the term \( e^{-\alpha r}/r \) diverges as $r\to 0$, it is convenient to rewrite it using the identity
\[
\frac{e^{-\alpha r}}{r} = \frac{e^{\alpha r}}{r} - \frac{2\sinh(\alpha r)}{r},
\]
where the second term remains finite at the origin, since \( \sinh(\alpha r)/r \to \alpha \) in the limit \( r \to 0 \). The divergent piece \( e^{\alpha r}/r \) is absorbed into the integral part of the general solution and therefore regularity at the origin requires that the coefficient of this divergent factor vanishes. This fixes the value of \( h_{00}(0) \) and yields a fully regular potential.
\\
Equivalently, the same regular solution can be obtained more directly by an appropriate change of integration limits, leading to the compact expression

\[
h_{00}(r) = -\frac{\beta e^{-\alpha r}}{r} \int_0^r e^{2\alpha u} \int_u^{\infty} e^{-\alpha p} \int_0^p \int_q^{\infty} s \rho(s) \, ds \, dq \, dp \, du,
\]

which is manifestly finite at \( r = 0 \). This form explicitly satisfies the physical requirements of asymptotic flatness and regularity at the origin, and is fully determined by the source $\rho(r)$. Finally, through successive integration by parts, we arrive at the explicit form of the metric perturbation
\begin{multline}\label{hI}
    h_{00}(r)= \frac{\beta}{\alpha^2}\left[-\frac{1}{r}\int_0^rs^2\rho(s)ds-\int_r^{\infty}s\rho(s)ds+\right.\\
    \left.\frac{\sinh(\alpha r)}{\alpha r}\int_r^{\infty}e^{-\alpha s}s\rho(s)ds+\frac{e^{-\alpha r}}{\alpha r}\int_0^r \sinh(\alpha s)s\rho(s)ds\right].
\end{multline}
The modified gravitational potential (\ref{hI}) can be naturally decomposed into four distinct contributions, each with a clear physical interpretation. The first two terms coincide with the standard Newtonian expressions for the interior and exterior mass contributions, respectively. The remaining two terms encode the Yukawa-type corrections inherent to \(f(R)\) gravity through exponentially weighted integrals, with the factors $\sinh(\alpha r)/(\alpha r)$, and $e^{\alpha r}$, which ensure that contributions from distant regions are strongly suppressed. As a result, the potential remains dominated by the local structure of the source, with nonlocal effects confined to a finite interaction range set by $\alpha^{-1}$.
\\
In the General Relativity limit (\(\alpha\to\infty\)), the exponential modifications vanish identically and the potential recovers the standard Newtonian form. The structure of Eq. (\ref{hI}) thus transparently separates the conventional mass contributions from the finite-range corrections induced by the quadratic curvature terms, highlighting how the radial matter distribution interacts with the screening scale \(\alpha^{-1}\).

\section{\label{sec:level4}Spherical Density Profiles and Modified Potentials}
In this section, we evaluate the modified gravitational potential for a variety of spherically symmetric mass distributions. These include idealized models such as the spherical shell and homogeneous sphere, astrophysically motivated profiles like Plummer, Hernquist, and NFW, as well as new density models introduced for comparison. Each profile is examined for its qualitative behavior and its implications under the modified gravity framework. For all cases we set $\beta=-8\pi  \alpha^2$.
\subsection{Spherical shell}
\begin{figure}[h!]
    \centering
    \includegraphics[width=0.6\textwidth]{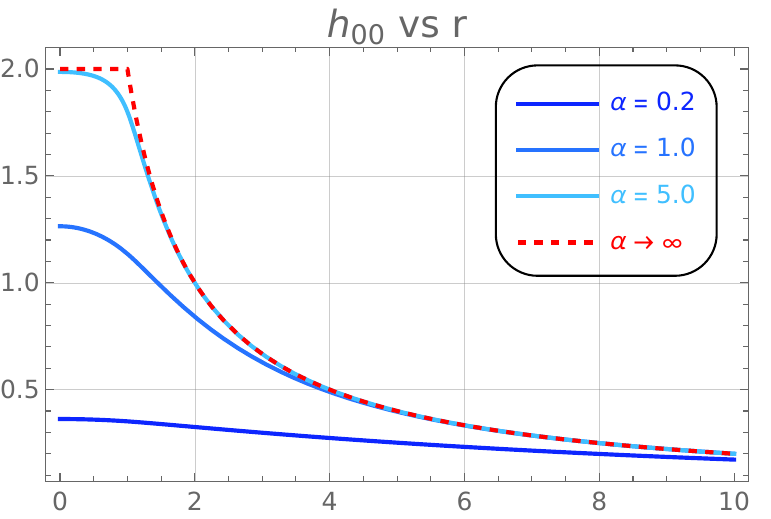}
    \caption{Modified gravitational potential $h_{00}(r)$ for a spherical shell of radius $R=1$ in units of $L$, for different values of the parameter $\alpha$ (in $L^{-1}$). The dashed curve represents the Newtonian limit ($\alpha\to\infty$), while solid curves correspond to finite $\alpha$, showing the transition from Yukawa-corrected to classical gravity. We set $M=1$ (in $L$).}
    \label{fig:01}
\end{figure}

We first consider an infinitesimally thin spherical shell of mass $M$ and radius $R$, described by the density distribution
\begin{equation}
    \rho(r)=\frac{M}{4\pi R^2}\delta(r-R),
\end{equation}
despite its idealized nature, this model provides a useful test case for exploring the behavior of the modified gravitational potential, which is given by 
\begin{equation}\label{solh2}
    h_{00}(r)=2M\left\{
    \begin{array}{cc}
    \left[\alpha r-e^{-\alpha R}\sinh (\alpha r)\right]\left(\alpha Rr\right)^{-1}\text{,} & r<R\\
    \left[\alpha R-e^{-\alpha r}\sinh(\alpha R)\right]\left(\alpha Rr\right)^{-1}\text{,} & r\geq R\\
    \end {array}
    \right.
\end{equation}
Inside the shell ($r<R$), the potential remains finite and exhibits a nontrivial radial dependence, unlike in Newtonian gravity where the potential is constant. 
When $r\to 0$, the potential takes the value
\begin{equation}
    h_{00}(0)=2M\frac{1-e^{-\alpha r_0}}{r_0},
\end{equation}
and at $r=R$ the potential is continuous. For $r>R$, the potential decays faster than $1/r$ due to the exponential suppression, reflecting the non-local nature of the Yukawa-type modification. When $\alpha\to\infty$ the exponential terms vanish and the standard Newtonian potential is recovered, as expected. The behavior of the potentials for some values of $\alpha$ are shown in Fig. \ref{fig:01}.
\subsection{Gaussian Mass Distribution}
We now turn our attention to a different configuration: a smooth, centrally concentrated and spherically symmetric mass distribution described by a Gaussian profile
\begin{equation}
    \rho(r)=\frac{M}{\left (\sqrt {\pi}\epsilon\right)^3}e^{-r^2/\epsilon^2},
\end{equation}
where $M$ represents the total mass and $\epsilon$ defines a characteristic spatial scale that determines the mass configuration, from point-like ($\epsilon\to 0$) to spatially extended ($\epsilon>0$) distribution. This model is particularly relevant when avoiding central singularities or modeling extended compact objects. The corresponding potential $h_{00}(r)$ is given by:
\begin{equation}
   h_{00}(r)=\frac{M}{r}\left\{
   2\text{E}\left(\frac{r}{\epsilon}\right)+
   e^{\left(\alpha\epsilon/2\right)^2-\alpha r}
   \left[
   \text{Ec}\left(\frac{r}{\epsilon}-\frac{\alpha\epsilon}{2}\right)+
   e^{2\alpha r}\text{Ec}\left(\frac{r}{\epsilon}+\frac{\alpha\epsilon}{2}\right)-2  \right]
   \right\}
\end{equation}
where 
\begin{equation}
    \text{E}(z)=\frac{2}{\sqrt{\pi}}\int_0^ze^{-t^2}\, dt,
\end{equation}
and $\text{Ec}(z)=1-\text{E}(z)$, denote the error function and complementary error function, respectively. The modified potential generated by the Gaussian mass distribution remains manifestly regular at the origin. Explicitly, the central value
\begin{equation}
    h_{00}(0)=2\alpha M e^{(\alpha \epsilon/2)^2} \text{Ec}\left(\frac{\alpha \epsilon}{2}\right),
\end{equation}
is finite for all values of the modified gravity parameter $\alpha$ and the width $\epsilon$. In the limit $\alpha\to\infty$, the potential reduces to
$2M\text{E}\left(r/\epsilon\right)/r$, which regularizes the Newtonian potential at short distances, remaining finite at the origin, and
\begin{equation}
    \lim_{\alpha\to\infty}h_{00}(0)=\frac{4M}{\sqrt{\pi}\epsilon},
\end{equation}
this result shows that the modified theory smoothly reproduces the classical behavior while preserving regularity at the origin.
\begin{figure}[h!]
    \centering
    \includegraphics[width=0.6\textwidth]{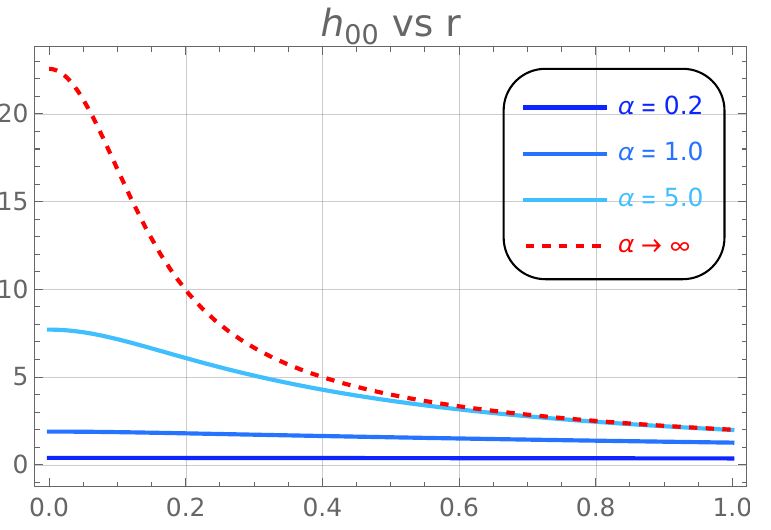}
    \caption{Modified gravitational potential $h_{00}(r)$ generated by a Gaussian mass distribution with width $\epsilon = 0.1$ (in units of $L$), shown for several values of the modified gravity parameter $\alpha$ (in $L^{-1}$). Solid curves correspond to finite $\alpha$, illustrating the Yukawa-type deviations induced by the quadratic $f(R)$ correction, while the dashed curve denotes the Newtonian limit $\alpha \to \infty$. The mass is fixed to $M = 1$ in units of $L$.}
    \label{fig:02}
\end{figure}
\subsection{Homogeneous Sphere}
\begin{figure}[h!]
    \centering
    \includegraphics[width=0.6\textwidth]{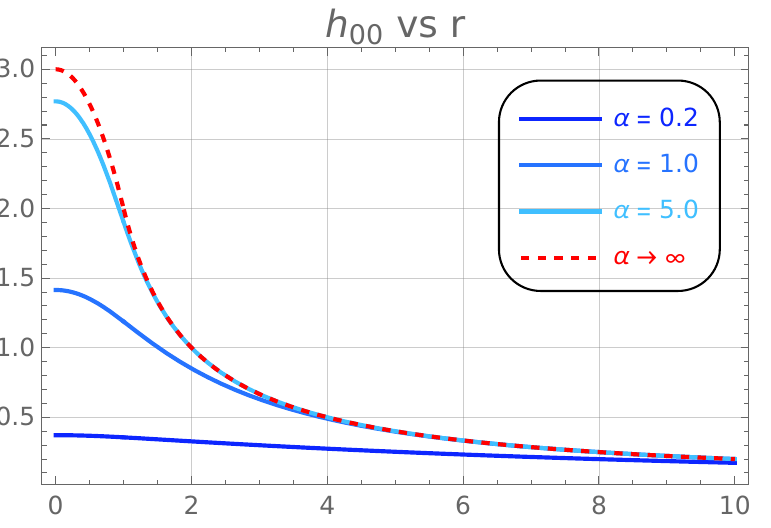}
    \caption{Radial behavior of the modified gravitational potential $h_{00}(r)$ generated by a homogeneous sphere of radius $R = 1$ (in units of $L$) for several values of the parameter $\alpha$ (in $L^{-1}$). Solid curves illustrate the impact of finite-$\alpha$ corrections to the Newtonian potential, while the dashed curve corresponds to the General Relativity limit $\alpha \to \infty$. The mass density is fixed to $\rho_0 = 1$ (in $L^{-2}$).}
    \label{fig:1}
\end{figure}

We now examine the case of a uniform density sphere, characterized by a constant mass density within a radius $R$, which is a natural approximation for stellar interiors and compact objects,
\begin{equation}\label{solh3}
    \rho(r)=\left\{
    \begin{array}{cc}
    \rho_0\text{,} & r<R\\
    0\text{,} & r\geq R\\
    \end {array}
    \right.
\end{equation}
the modified potential for such a configuration splits into two regimes, interior ($r<R$), combining exponential corrections with polynomial terms, 
\begin{equation}\label{symmetry}
    h_{00}(r)=\frac{4\pi\rho_0}{\alpha^2}\left[\frac{e^{-\alpha(r+R)}}{r}\left(e^{2\alpha r}-1\right)\left(\frac{1}{\alpha}+R\right)+\alpha^2\left(\frac{r^2}{3}-R^2\right)-2\right],
\end{equation}
with finite value at origin
\begin{equation}
    h_{00}=\frac{6M}{\alpha^2 R^3}\left[e^{-\alpha R}(1+\alpha R)+\frac{\alpha^2R^2}{2}-1\right];
\end{equation}
and exterior ($r>R$), where the potential takes a Yukawa-type form, with exponential suppression at large distances
\begin{equation}\label{integral}
    h_{00}(r)=\frac{8\pi\rho_0}{3\alpha^3r}\left[\left(\alpha R\right)^3+3e^{-\alpha r}\left(\sinh(\alpha R)-\alpha R\cosh(\alpha R)\right)\right].
\end{equation}
In both cases the expression recovers the classical result in the Newtonian limit
$4\pi\rho_0\left(R^2-r^2/3\right)$ if $r<R$, and $2M/r$, where $M=4\pi R^3\rho_0$, when $R<r$. This can be seen in Fig. \ref{fig:1}, where $h_{00}(r)$ is plotted as a function of radial distance $r$, for different values of the $f(R)$ parameter $\alpha$. This qualitative pattern, regular behavior in the interior and convergence to the Newtonian profile in the large $\alpha$ limit, is common to all the following mass distributions analyzed in this work.

\subsection{Plummer Density Profile}
\begin{figure}[h!]
    \centering
    \includegraphics[width=0.6\textwidth]{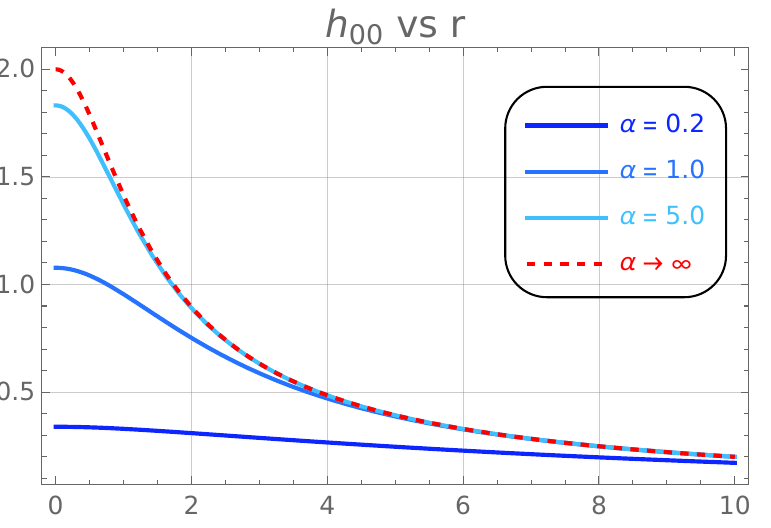}
    \caption{Modified gravitational potential $h_{00}(r)$ for the Plummer model, computed numerically varying $\alpha$. The dashed curve represents the Newtonian potential ($\alpha\to\infty$), while solid curves show $f(R)$-corrected profiles. $M=1$ (in $L$), $b=1$ (in $L$).}
    \label{fig:2}
\end{figure}

The Plummer model is a widely used smooth and finite-density profile in astrophysics \cite{Plummer}, especially for modeling globular clusters and spherical stellar systems. It features a core-like structure with a central density plateau and a gradual fall-off at large radii. The mass density is given by
\begin{equation}
    \rho(r)=\frac{3 r_s^2 M}{4\pi\left(r_s^2+r^2\right)^{5/2}},
\end{equation}
where $M$ is the total mass and $r_s$ is a scale radius. While the complexity of this density distribution prevents a fully analytical solution for $h_{00}(r)$ in $f(R)$ gravity, it is possible to obtain the value of the potential at $r=0$, which takes the finite value
\begin{equation}
    h_{00}(0)=\pi\alpha M\left[\frac{2}{\pi b}+b\left(\textbf{H}_{-3}(b)-Y_1(b)\right)+4\textbf{H}_{-2}(b)\right],
\end{equation}
where $b=\alpha r_s$, $\textbf{H}_n(x)$ is the Struve function, and $Y_n(x)$ is the Bessel function of the second kind. We compute the potential numerically as a function of the parameter $\alpha$, as shown in Fig. \ref{fig:2}, where it can be seen that as $\alpha$ increases, the potential converges towards the Newtonian limit (dashed curve), $2M/\sqrt{b^2+r^2}$.

\subsection{Hernquist Profile}
\begin{figure}[h!]
    \centering
    \includegraphics[width=0.6\textwidth]{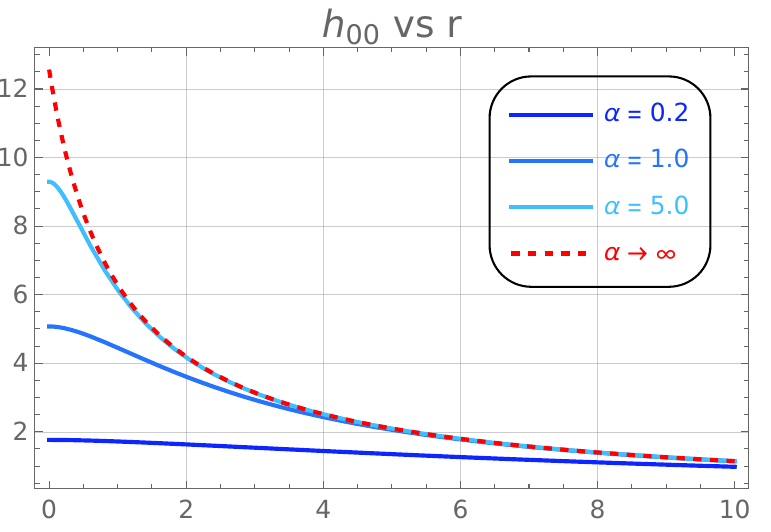}
    \caption{Modified potential $h_{00}(r)$ for the Hernquist density, for some values of $\alpha$. The dashed curve represents the Newtonian potential. $\rho_0=1$ (in $L$) and $r_s=1$ (in $L$).}
    \label{fig:4}
\end{figure}

The Hernquist profile is a widely used model to describe the mass distribution in elliptical galaxies and bulges of spiral galaxies \cite{hernquist}. It features a cuspy central density and a steep fall-off at large radii. Its density is given by
\begin{equation}\label{models}
    \rho(r)=\frac{\rho_0 r_s^{\mu}}{r^{\mu}\left(1+r/r_s\right)^{\nu-\mu}}
\end{equation}
with parameters $\mu=1$ and $\nu=4$, and $r_s$ representing a scale radius. For this profile, the modified gravitational potential was computed as
\begin{multline}\label{tracesolution}
    h_{00}(r)=\frac{4\pi \rho_0 r_s^3}{r}\left(1-e^{\alpha r}-b e^{b}\sinh(\alpha r)\,\text{E}(b+\alpha r)+\right.\\
    \left. b e^{-\alpha r}\left[\sinh{b}\,\text{Chi}(b+\alpha r)-\cosh{b}\, \text{Shi}(b+\alpha r)+c(b)\right]\right),
\end{multline}
where $b=\alpha r_s$, $\text{Chi}(x)$ and $\text{Shi}(x)$ are the hyperbolic cosine and sine integral functions, respectively. $\text{E}(x)$ is the exponential integral function, defined by
\begin{equation}
    \text {E} (x) = \int_ {1}^\infty\frac {e^{-x t}} {t}\, dt,
\end{equation}
and the constant, determined by the parameters $\alpha$ and $r_s$, is
\begin{equation}
    c(x)=\cosh{x}\, \text{Shi}\,x-\sinh{x}\,\text{Chi}\,x
\end{equation}
Solution (\ref{tracesolution}) reveals the regularity at $r=0$, since
\begin{equation}
    h_{00}(0)=4\pi \rho_0\, r_s^2\, b\left[1-b\, e^{b}\, \text{E}(b)\right].
\end{equation}
Likewise we can observe two key features in the potential: (i) a dominant Newtonian-like term $\propto r^{-1}$, and (ii) $f(R)$ induced corrections governed by $\text{E}(x)$. For large $\alpha$ the potential can be written as
\begin{equation}
    h_{00}(r)\approx\frac{4 \rho_0 r_s \pi}{\alpha^2r}\left[2e^{-\alpha r}+\frac{r_s^2}{r_s+r}\left(\alpha^2 r-\frac{2r_s}{(r_s+r)^2}\right)\right],
\end{equation}
therefore the classical Hernquist potential, $4r_s^3\pi\rho_0/(r_s+r)$, is recovered when $\alpha\to\infty$, while finite $\alpha$ values exhibit Yukawa screening corrections. In Fig. \ref{fig:4} the behavior of the potential (\ref{tracesolution}) is depicted for some numerical values of the parameters. Unlike the previously analyzed profiles, the Hernquist potential does not exhibit a smooth, concave-down behavior as $r\to0$, this is a direct consequence of the cuspy nature of the Hernquist density profile, $\rho(r)\propto r^{-1}$ at small radii. Although the modified gravity framework introduces Yukawa-type corrections that soften the interaction at finite scales, the underlying central divergence of the mass density persists and is reflected in $h_{00}(r)$. As a result, the potential develops a sharper radial dependence near the origin, lacking the regular flattening observed for cored distributions such as the homogeneous or Gaussian models. This behavior highlights the sensitivity of the modified potential to the inner slope of the mass profile, even within the linearized regime.
\subsection{Navarro–Frenk–White (NFW) Profile}
\begin{figure}[h!]
    \centering
    \includegraphics[width=0.6\textwidth]{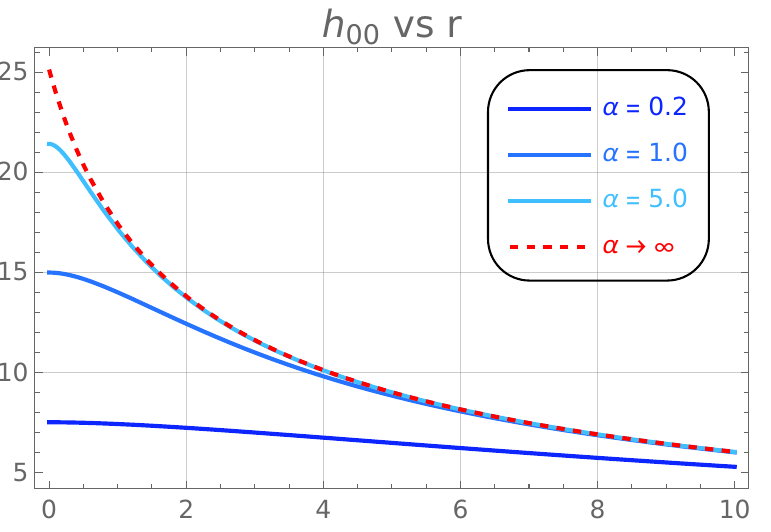}
    \caption{Modified gravitational potential $h_{00}(r)$ for some values of the parameter $\alpha$ for the NFW density profile. The piecewise curve represents the Newtonian limit. Compared to the Hernquist case, the deviation between the modified and classical potentials is more pronounced at small radii, reflecting the cuspy nature of the NFW profile. In this plot $\rho_0=1$ (in $L$) and $r_s=1$ (in $L$).}
    \label{fig:5}
\end{figure}

We now analyze the NFW profile \cite{NFW}, which is given by Ec. (\ref{models}) with $\mu=1$ and $\nu=3$, i.e.,
\begin{equation}
    \rho(r)=\frac{r_s\rho_0}{r(1+r/r_s)^2}.
\end{equation}
This model is a cornerstone of modern cosmology, emerging from N-body simulations of cold dark matter halos, because it provides a universal, simulation-derived description of dark matter halo density \cite{Wyithe_2001,Lokas_2001,Kuzio_de_Naray_2009,Dutton_2014}, bridging theoretical predictions of hierarchical structure formation with observational data. It describes a density distribution with a cuspy center and a slower decay ($\rho\propto r^{-3}$) at large radii and therefore modifies the potential’s asymptotic behavior relative to the Hernquist case. We derive the modified potential
\begin{multline}\label{wavemodi}
    h_{00}(r)=\frac{8\pi r_s^3}{r}\left[\ln\left(1+r/r_s\right)-e^b\text{Ei}(-b-\alpha r)\sinh(\alpha r)\right.\\
    \left.-e^{-\alpha r}\left(\cosh{b}\,\text{Chi}(b+\alpha r)-\sinh{b}\,\text{Shi}(b+\alpha r)+k(b)\right)\right],
\end{multline}
where again $b=\alpha r_s$ and and $\text{Ei}(x)$ is the exponential integral function
\begin{equation}
    \text{Ei}(x)=\int_{-\infty}^x\frac{e^t}{t}\, dt,
\end{equation}
and
\begin{equation}
    k(x)=\sinh{x}\, \text{Shi}\,b-\cosh{b}\, \text{Chi}\, b,
\end{equation}
and $s=\alpha(r_s+r)$. The logarithmic term reflects the long-range behavior typical of NFW-like profiles, while the exponential integrals introduce scale-dependent modifications governed by $\alpha$. In Fig. \ref{fig:5} the potential (\ref{wavemodi}) is displayed for some numerical values of the parameters, allowing a clear visualization of the deviations from the classical case across the radial range.
As in the previous models, the potential is finite at the origin
\begin{equation}
    h_{00}(0)=-8\pi\rho_0\, r_s^2\, b\, e^{b} \text{Ei}(-b),
\end{equation}
and as $\alpha$ increases, the potential takes the form
\begin{equation}\label{tmodif}
    h_{00}(r)\approx \frac{8\pi\rho_0 r_s^3}{\alpha^2r}\left[\frac{e^{-\alpha r}}{r_s^2}+a^2\ln\left(1+\frac{r}{r_s}\right)-\frac{1}{(r_s+r)^2}\right],
\end{equation}
where the exponential corrections are suppressed, and the potential approaches its standard Newtonian form, $8r_s^3\pi \ln(1+r/r_s)/r$. Conversely, for small $\alpha$, the deviations become more pronounced, especially at large radii, which could have observable consequences in galactic halo dynamics under extended gravity theories.


\subsection{Exponential Cutoff Profile}
\begin{figure}[h!]
    \centering
    \includegraphics[width=0.6\textwidth]{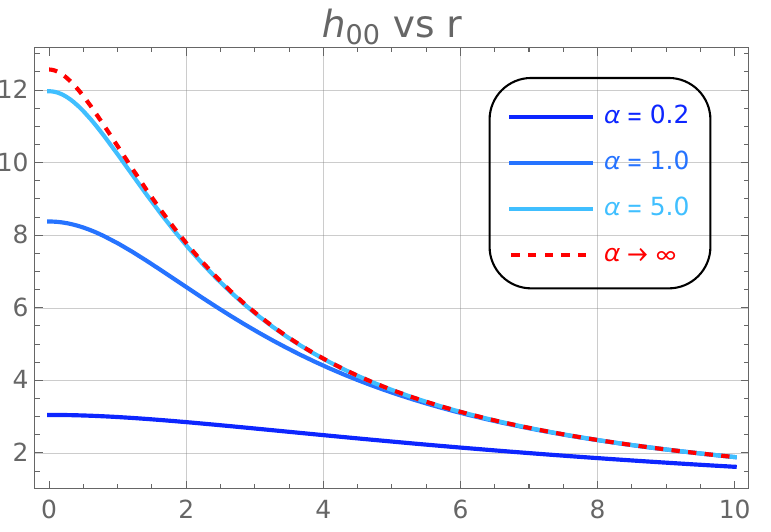}
    \caption{Modified gravitational potential for the exponential cutoff profile Eq. (\ref{inven1}), plotted for $\lambda=1$ (in $L^{-1}$) and $\rho_0=1$ (in $L^{-2}$) and varying $\alpha$.}
    \label{fig:3}
\end{figure}

We propose a new spherical density profile with exponential suppression at large distances and a regular behavior near the origin,
\begin{equation}\label{inven1}
    \rho(r)=\frac{\rho_0}{\lambda}\frac{1-e^{-\lambda r}}{r}e^{-\lambda r}.
\end{equation}
where the parameter $\lambda$ is an inverse length scale that controls the spatial extent and decay of the density. This profile smoothly decreases from $\rho_0$ and asymptotically tends to zero at large radii due to the exponential cutoff $e^{-\lambda r}$, guaranteeing a finite total mass. This form could be well-suited to modeling astrophysical systems characterized by a soft core and a rapidly declining outer density, including dark matter halos with cored density distributions and globular clusters. The corresponding potential is
\begin{equation}\label{tensorpert}
    h_{00}(r)=\frac{6\pi\rho_0\left(1-e^{-\lambda r}\right)}{\lambda^3(1-5b^2+4b^4)r}\left(\frac{b^2-1}{3e^{\lambda r}}+4b^4\frac{1-e^{-\alpha r}}{1-e^{-\lambda r}}-5b^2+1\right),
\end{equation}
where $b=\lambda/\alpha$. 
This function exhibits continuity for all values of $\alpha$ and $r$, this can be seen Fig. \ref{fig:3}, where the potential is shown as a function of $r$ for some values of $\alpha$. At the origin, the potential takes the value
\begin{equation}
    h_{00}(0)=\frac{4\pi\rho_0\, \alpha(\alpha+3\lambda)}{\lambda^2(\alpha+\lambda)(\alpha+2\lambda)}.
\end{equation}
For the special case $\alpha=\lambda$ and $\alpha=2\lambda$, where the potential denominator vanishes, the solution remains finite and smooth, given by
\begin{equation}
    \lim_{\alpha\to\lambda}h_{00}(r)=\frac{2\pi\rho_0}{3\lambda^3r}\left[9-e^{-2\lambda\, r}-2e^{-\lambda\, r}\left(4+3 \lambda\,r\right)\right],
\end{equation}
and
\begin{equation}
    \lim_{\alpha\to2\lambda}h_{00}(r)=\frac{2\pi\rho_0}{3\lambda^3r}\left[9-16e^{-\lambda\, r}+e^{-2\lambda\, r}\left(7+3 \lambda\,r\right)\right],
\end{equation}
respectively. Most notably, the Newtonian limit 
\begin{equation}
    \lim_{\alpha\to\infty}h_{00}(r)=\frac{2\pi\rho_0}{\lambda^3r}\left(3+e^{-2\lambda r}-4e^{-\lambda r}\right),
\end{equation}
yields a finite potential at the origin, 
\begin{equation}
    \lim_{r\to0}\lim_{\alpha\to\infty}h_{00}(r)=\frac{4\pi\rho_0}{\lambda^2}.
\end{equation}
ensuring regularity in this scenario.
\subsection{Linear-Exponential Profile}
Another smooth model is the linear–exponential profile
\begin{equation}\label{quadrupole}
    \rho(r)=\rho_0re^{-\lambda r}.
\end{equation}
This model grows linearly near the origin, reaching a maximum before decaying exponentially, it could serve to model matter distributions with vanishing central density and a peak at a finite radius. The corresponding modified gravitational potential is given by
\begin{multline}\label{termcorrection}
    h_{00}(r)=\frac{48\pi\rho_0e^{-s}}{\lambda^2}\left(\frac{e^{s}}{s}-\frac{s}{6\left(1-b^2\right)}-\frac{2(1-2b^2)}{3(1-b^2)}\right.\\\left.-\frac{3(1-3b^2)+b^4\left[10-(1+3b^2)e^{s-s/b}\right]}{3(1-b^2)^3s}\right),
\end{multline}
where $b=\lambda/\alpha$ and $s=\lambda r$. This potential remains continuous for all values of $\alpha$, since for $\alpha=\lambda$, we have
\begin{equation}
    \lim_{\alpha\to\lambda}h_{00}(r)=\frac{2\pi\rho_0}{3\lambda^2s}\left[72-e^{-s}(3+s)\left(24+s(9+2s)\right)\right],
\end{equation}
similarly, the potential is regular at the origin
\begin{equation}
    h_{00}(0)=16\pi\rho_0\lambda\left(\frac{1}{\lambda^3}-\frac{1}{(\alpha+\lambda)^3}\right),
\end{equation}
and the Newtonian limit
\begin{equation}\label{correction}
    \lim_{\alpha\to\infty}h_{00}(r)=\frac{8\pi\rho_0}{\lambda^2 s}\left(6-e^{-s}(6+4s+s^2)\right).
\end{equation}
which is regular at the origin, with
\begin{equation}
    \lim_{r\to0}\lim_{\alpha\to\infty}h_{00}(r) = \frac{16\pi\rho_0}{\lambda^2}.
\end{equation}
This behavior reinforces the consistency of the model and suggests its suitability for describing finite, centrally depleted mass configurations in modified gravity frameworks.
\\
It is worth to note that both models, (\ref{inven1}) and (\ref{quadrupole}), share Yukawa like decay ($\sim e^{-\alpha r}/r$) for $r\gg \lambda^{-1}$, but the linear-exponential profile exhibits stronger suppression at large $r$ due to the additional power law factor in $\rho(r)$.

\subsection{Exponential-Singular Profile}
As another toy model, we consider the density profile
\begin{equation}\label{inven3}
    \rho(r)=\rho_0\frac{e^{-\lambda r}}{\lambda\, r},
\end{equation}
which introduces a singularity in the mass density as $r\to0$ when $\lambda>0$, softened by the exponential decay. For this distribution, the modified potential takes the form
\begin{equation}
    h_{00}(r)=\frac{8\pi\rho_0}{\lambda^3(\alpha^2-\lambda^2) r}\left[\alpha^2\left(1-e^{-\lambda r}\right)-\lambda^2\left(1-e^{-\alpha r}\right)\right],
\end{equation}
which remains continuous for all $\alpha$, in particular for $\alpha=\lambda$
\begin{equation}
    \lim_{\alpha\to\lambda}=\frac{4\pi\rho_0}{\lambda^3r}\left[2-(2+\lambda\, r)e^{-\lambda\, r}\right],
\end{equation}
moreover, the potential is finite at origin 
\begin{equation}
    h_{00}(0)=\frac{8\pi\rho_0\alpha}{\lambda^2(\alpha+\lambda)},
\end{equation}
and in the Newtonian limit the potential simplifies to
\begin{equation}
    \lim_{\alpha\to \infty}h_{00}(r)=\frac{8\pi\rho_0\left(1-e^{-\lambda r}\right)} {\lambda^3 r},
\end{equation}
this behavior indicates that the model yields a regular potential in the Newtonian case, $8\pi\rho_0/\lambda^2$, despite the divergence in the density at small $r$. The profile may be useful for approximating mass distributions with steep inner density rises but finite total mass, and it provides analytic insight into the interplay between scale parameters in extended gravity models.

\section{\label{sec:level5}Rotation Curves}
\begin{figure}[h!]
    \centering
    \includegraphics[width=0.6\textwidth]{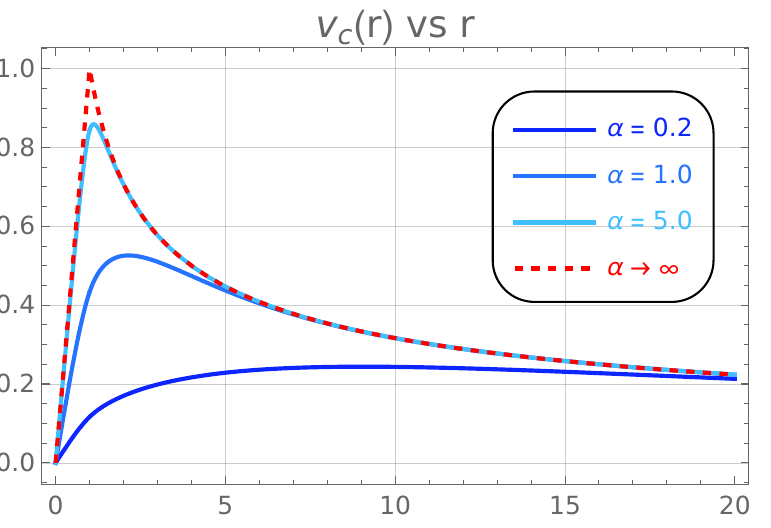}
    \caption{Rotation curve $v_c(r)$ (dimensionless) derived from the modified gravitational potential of a homogeneous sphere for several values of the quadratic gravity parameter, $\alpha$. In the Newtonian ($\alpha \to \infty$, dashed line), the rotation curve exhibits a pronounced maximum; however, finite values of $\alpha$ smooth the inner rise of the rotation curve and enhance the velocity at intermediate radii, illustrating the scale-dependent modifications induced by the quadratic $f(R)$ term. In this plot $M=1$ in units of $L$.}
    \label{fig:9}
\end{figure}

\begin{figure}[h!]
    \centering
    \includegraphics[width=0.6\textwidth]{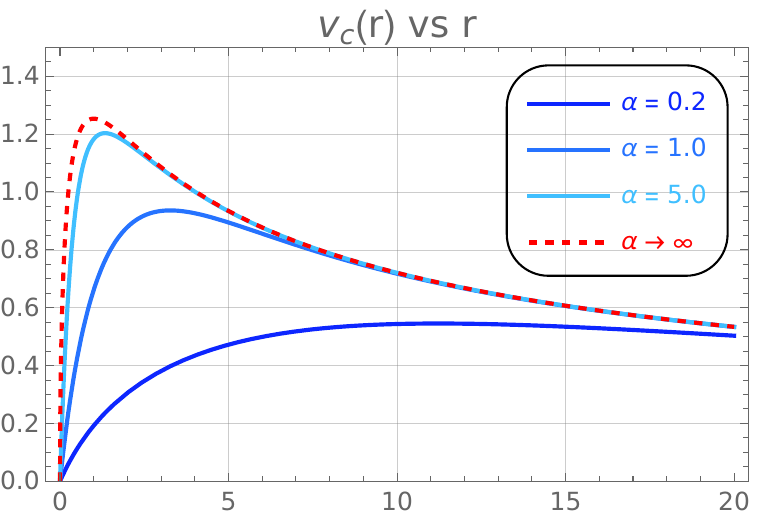}
    \caption{Rotation curves $v_c(r)$ obtained from the modified gravitational potential associated with the Hernquist density profile, for several values of the parameter $\alpha$. In contrast with the homogeneous sphere, the Hernquist model does not exhibit a pronounced central peak in the Newtonian regime, reflecting the cuspy nature of the underlying density profile and leading to a smoother rise of the rotation velocity near the galactic center. $\rho_0=1$ (in $L$) and $r_s=1$ (in $L$)}
    \label{fig:10}
\end{figure}

\begin{figure}[h!]
    \centering
    \includegraphics[width=0.6\textwidth]{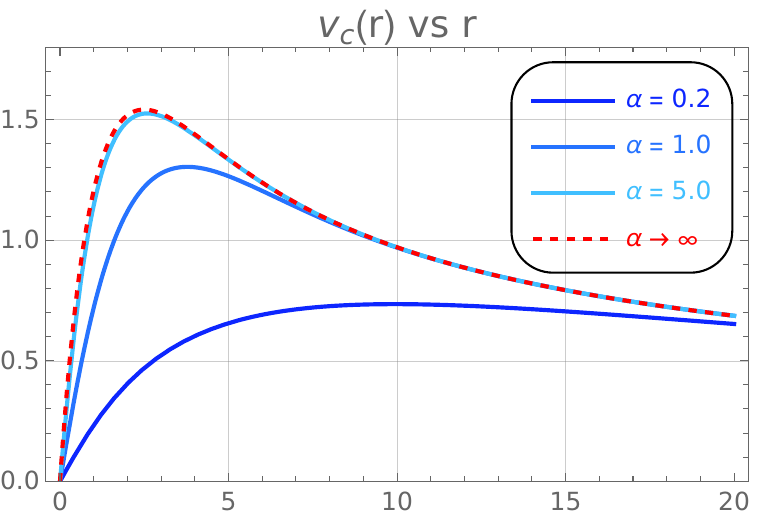}
    \caption{Rotation curves $v_c(r)$ derived from the modified gravitational potential for the exponential cutoff density profile, Eq. (\ref{inven1}), shown for different values of $\alpha$. Compared to the Hernquist case, the maximum of the rotation curve is shifted to slightly larger radii, reflecting the softer central concentration of the mass distribution. We set $\lambda=1$ (in $L^{-1}$) and $\rho_0=1$ (in $L^{-2}$).}
    \label{fig:11}
\end{figure}

In this section, we compute the galactic rotation curves predicted by the modified gravitational potentials of the homogeneous sphere, Hernquist, and exponential cutoff profiles, and compare them with observational data from the well-measured galaxy NGC 3198. We employ the high‑quality rotation curve and photometric data from the SPARC (Spitzer Photometry and Accurate Rotation Curves) database \cite{Lelli2016SPARC}.
\\
For each of the selected density profiles, we compute the corresponding rotation velocity within the linearized quadratic $f(R)$ framework. The resulting theoretical rotation curves are presented in Figs. \ref{fig:9}, \ref{fig:10} and \ref{fig:11}.
Since the parameter $\alpha$ effectively suppresses the central peak of the rotation curve and yields a smoother transition at intermediate and large galactic radii, the free parameters of each density profile can be tuned to improve agreement with observational data. To assess this quantitatively, we performed a $\chi^{2}$ minimization to determine the optimal values of both the structural parameters of each model and the screening scale $\alpha$.
\\
It should be noted, however, that the gravitational potentials derived in this work were constructed to satisfy physical boundary conditions, namely regularity at the origin and asymptotic flatness, which imply that the potential vanishes at spatial infinity. As a consequence, the resulting rotation curves cannot reproduce a strictly flat behavior at arbitrarily large radii. For this reason, the fitting procedure was restricted to the inner galactic region, specifically to radii $r \lesssim 30$ $\mathrm{kpc}$, where the linearized quadratic $f(R)$ framework is expected to provide a reliable description. The corresponding best fit parameters and reduced chi-squared values are summarized in Table~\ref{tab:chi2_fits}. This comparison is not intended as a precision fit, but as a quantitative illustration of the level of agreement achievable within the linearized quadratic $f(R)$ framework.

\begin{figure}[h!]
    \centering
    \includegraphics[width=0.7\textwidth]{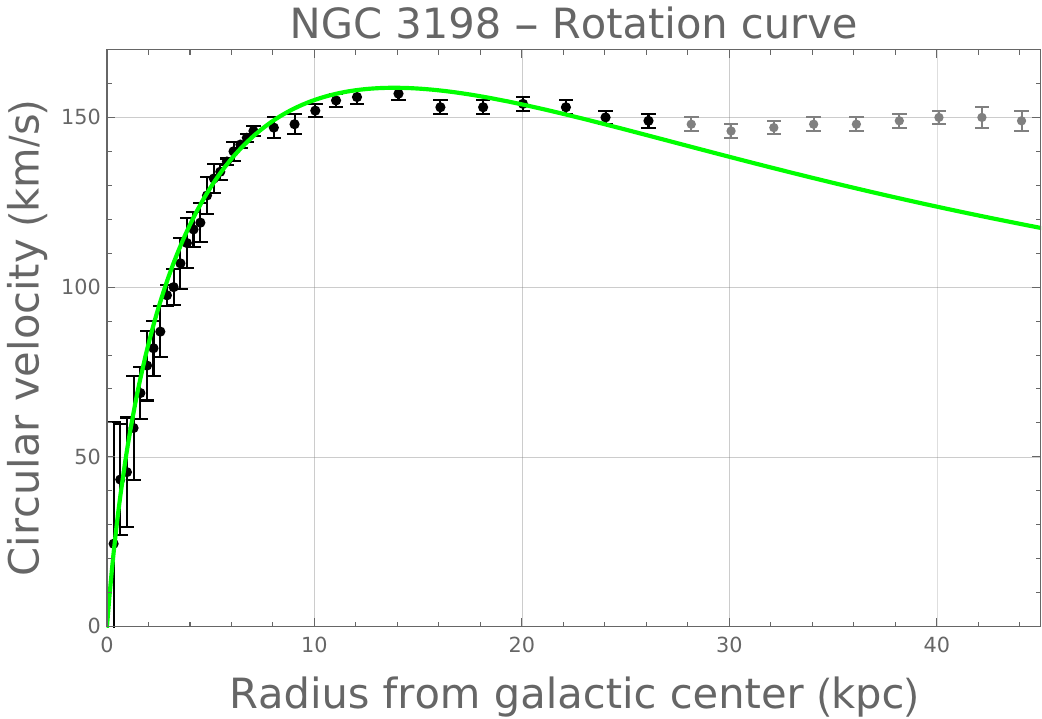}
    \caption{Comparison between the observed rotation curve of NGC 3198 and the circular velocity derived from the Hernquist potential in linearized quadratic $f(R)$ gravity, Eq. (\ref{tracesolution}). Black data points correspond to the observed velocities included in the $\chi^2$ minimization, while gray points represent observational data excluded from the fit. The solid green curve shows the best-fit theoretical model. The agreement is satisfactory in the inner and intermediate radial range; however, for radii larger than $\sim 30$ kpc the model fails to reproduce the observed nearly flat behavior, as the potential asymptotically decays to zero at large distances.}
    \label{fig:12}
\end{figure}

\begin{table}[ht]
\centering
\renewcommand{\arraystretch}{1.2}
\begin{tabular}{l l c}
\hline\hline
Model & Best fit parameters & $\chi^2_{\mathrm{red}}$ \\
\hline
Homogeneous sphere
& $M = 6.52\times10^{5},\ \alpha = 0.13,\ R = 2.52$ 
& 0.961 \\

\hspace{20pt}(Newtonian limit) 
& Not applicable (pronounced central peak) 
& -- \\

Hernquist
& $\rho_0 = 3.30\times10^{5},\ \alpha = 0.148,\ r_s = 0.679$ 
& 1.211 \\

\hspace{20pt}(Newtonian limit) 
& $\rho_0 = 59.1,\ r_s = 16.1$ 
& 2.821 \\

Exponential cutoff
& $\rho_0 = 3.62\times10^{5},\ \alpha = 0.128,\ \lambda = 1.73$ 
& 1.129 \\

\hspace{20pt}(Newtonian limit) 
& $\rho_0 = 4.31\times10^{2},\ \lambda = 0.197$ 
& 4.163 \\
\hline\hline
\end{tabular}
\caption{Best fit parameters and reduced chi-square values obtained from rotation curve fits to SPARC data for the galaxy NGC 3198, using only radial data up to $r\leq 30$ kpc. Results are shown for different spherically symmetric mass models within linearized quadratic $f(R)$ gravity and in the Newtonian (GR) limit. Radial distances are expressed in kiloparsecs (kpc) and circular velocities in km/s; model parameters are reported in the corresponding physical units consistent with this choice. The quadratic $f(R)$ framework systematically yields improved fits, particularly for the homogeneous sphere, for which the Newtonian model fails due to the presence of a pronounced central peak in the rotation curve.}
\label{tab:chi2_fits}
\end{table}
\noindent Figure \ref{fig:12} shows the observed rotation curve of the galaxy NGC 3198 obtained from the SPARC database, together with the best-fit Hernquist model within linearized quadratic $f(R)$ gravity. The theoretical curve is computed using the optimal parameters reported in Table \ref{tab:chi2_fits}, obtained from the reduced chi-square minimization. The comparison illustrates the ability of the modified potential to reproduce the observed rotational behavior over the fitted radial range.

\section{\label{sec:level6}Conclusions}
In this work, the gravitational potential was analyzed within the framework of $f(R)$ gravity, considering a quadratic form of the function $f(R)$. The resulting potential satisfies a fourth-order differential equation (\ref{poisson}), which was solved analytically under the assumption of spherical symmetry. This equation exhibits two key features: (i) For $\alpha^2\to\infty$ and finite $\gamma$, the $\nabla^4$ term is suppressed, reducing to the standard Poisson equation; and (ii) Yukawa-type corrections from the $\nabla^4$ operator introduces massive modes generating exponentially screened potentials.
\\
By imposing physical boundary conditions (regularity at the origin and asymptotic flatness) all integration constants were shown to vanish. This guarantees that the potential, Eq. (\ref{hI}), is uniquely determined by the mass distribution that sources the field.
\\
The solution for the modified gravitational potential, Eq. (\ref{hI}), was found to consist of the standard GR term supplemented by Yukawa-type corrections of the form $e^{-\alpha r}/r$, this term encodes the additional scalar interaction mediated by the scalaron field, with $\alpha^{-1}$ setting the screening scale. Accordingly, in the limit $\alpha\to\infty$, the potential smoothly reduces to the Newtonian form, fully recovering GR. Similarly, at large distances $r\gg\alpha^{-1}$, the potential exhibits classical behavior.
\\
The potential was examined for a range of spherically symmetric mass distributions, including idealized configurations, astrophysically motivated profiles, and proposed novel density functions. In all cases, the solutions were found to be continuous and smooth throughout the entire domain. The boundary conditions consistently led to vanishing integration constants, confirming that the potential is entirely determined by the matter distribution. While the Newtonian limit is recovered as $\alpha\to\infty$, finite values of $\alpha$ introduce nontrivial corrections at intermediate scales $r\sim\alpha^{-1}$, which could have important phenomenological implications in gravitational systems.
\\
Our analysis highlights a key structural property of the modified gravitational potentials in quadratic $f(R)$ gravity. When the integration constants are fixed by imposing regularity at the origin and asymptotic flatness, the resulting potentials remain finite at $r=0$ for all values of $\alpha$. In the GR limit $\alpha \to \infty$, these solutions smoothly reduce to their Newtonian counterparts.
\\
The radial behavior of the corrections depends on the underlying mass distribution. For diffuse and centrally regular profiles, such as the Plummer sphere, the deviations from the Newtonian potential are most significant at intermediate radii, where the Yukawa-like contributions compete with the classical term. In contrast, for cuspy density profiles like Hernquist and NFW, whose densities diverge as $r \to 0$, the modified potential inherits a steeper inner behavior, leading to more pronounced departures at small radii. For compact configurations, including spherical shells and exponentially decaying profiles, the corrections are spatially localized, becoming relevant primarily near the origin or close to the characteristic scale of the mass distribution.
\\
\\
The novel density profiles introduced in this work, characterized by simple analytical forms and tunable parameters such as decay rates or core slopes, offer alternatives for modeling astrophysical systems. As an illustrative example, we compare the circular velocity derived from the Hernquist profile in quadratic $f(R)$ gravity with the observed rotation curve of NGC 3198, by performing a simple $\chi^2$ analysis to quantify the agreement between the predicted circular velocity and the observed rotation curve. The resulting reduced chi-square indicates a moderate but not optimal fit, consistent with the expected limitations of the quadratic $f(R)$ model at large radii.
\\
We find that the modified gravitational potential provides a reasonable description of the observed rotation curve in the inner and intermediate radial range $r \lesssim 30$ kpc, and in several cases yields improved fits relative to the Newtonian limit for the same mass profiles. At larger radii, however, the predicted velocities systematically decline, reflecting the Yukawa-type suppression inherent to the linearized quadratic $f(R)$ model.
\\
The linearized quadratic $f(R)$ model does not generate asymptotically flat rotation curves, but it constitutes a controlled and predictive modification of Newtonian gravity at short and intermediate distances. The explicit analytical solutions derived in this work therefore provide a useful theoretical framework for testing deviations from General Relativity in compact astrophysical systems and in observational regimes where modified gravity effects may be constrained.

Although this work has established the mathematical framework for the gravitational potential solutions, we emphasize that our focus has been on their theoretical properties rather than observational detection. Future research could extend this study by exploring numerical simulations and strong gravity regimes (considered the most promising laboratories) where shallow potentials enhance the effects of modified gravity.

\nocite{*}
\section{References}
\bibliographystyle{vancouver}
\bibliography{sn-bibliography}

\providecommand{\noopsort}[1]{}\providecommand{\singleletter}[1]{#1}%
\begin{thebibliography}{10}

\bibitem{Sotiriou_2006}
Sotiriou TP.
\newblock f(R) gravity and scalar–tensor theory.
\newblock Classical and Quantum Gravity. 2006 aug;23(17):5117.
\newblock Available from: \url{https://dx.doi.org/10.1088/0264-9381/23/17/003}.

\bibitem{Sotiriou:2008rp}
Sotiriou TP, Faraoni V.
\newblock {$f(R)$ Theories Of Gravity}.
\newblock Rev Mod Phys. 2010;82:451-97.

\bibitem{De_Felice_2010}
De~Felice A, Tsujikawa S.
\newblock f(R) Theories.
\newblock Living Reviews in Relativity. 2010 Jun;13(1).
\newblock Available from: \url{http://dx.doi.org/10.12942/lrr-2010-3}.

\bibitem{Clifton_2012}
Clifton T, Ferreira PG, Padilla A, Skordis C.
\newblock Modified gravity and cosmology.
\newblock Physics Reports. 2012 Mar;513(1–3):1–189.
\newblock Available from: \url{http://dx.doi.org/10.1016/j.physrep.2012.01.001}.

\bibitem{Shankaranarayanan_2022}
Shankaranarayanan S, Johnson JP.
\newblock Modified theories of gravity: Why, how and what?
\newblock General Relativity and Gravitation. 2022 May;54(5).
\newblock Available from: \url{http://dx.doi.org/10.1007/s10714-022-02927-2}.

\bibitem{Hurtado_2023}
Hurtado R, Arenas R.
\newblock Hypergeometric viable models in $f(R)$ gravity.
\newblock Physica Scripta. 2023 jul;98(8):085001.
\newblock Available from: \url{https://dx.doi.org/10.1088/1402-4896/ace0e3}.

\bibitem{hurtadoRo}
Hurtado RA, Arenas R.
\newblock Spherically symmetric and static solutions in $f(R)$ gravity coupled with electromagnetic fields.
\newblock Phys Rev D. 2020 Nov;102:104019.
\newblock Available from: \url{https://link.aps.org/doi/10.1103/PhysRevD.102.104019}.

\bibitem{Capozziello:2004km}
Capozziello S, Cardone VF, Carloni S, Troisi A.
\newblock {Higher order curvature theories of gravity matched with observations: A Bridge between dark energy and dark matter problems}.
\newblock AIP Conf Proc. 2005;751:54-63.
\newblock [,54(2004)].

\bibitem{Ivanov_2022}
Ivanov VR, Ketov SV, Pozdeeva EO, Vernov SY.
\newblock Analytic extensions of Starobinsky model of inflation.
\newblock Journal of Cosmology and Astroparticle Physics. 2022 Mar;2022(03):058.
\newblock Available from: \url{http://dx.doi.org/10.1088/1475-7516/2022/03/058}.

\bibitem{Vilenkin}
Vilenkin A.
\newblock Classical and quantum cosmology of the Starobinsky inflationary model.
\newblock Phys Rev D. 1985 Nov;32:2511-21.
\newblock Available from: \url{https://link.aps.org/doi/10.1103/PhysRevD.32.2511}.

\bibitem{Nojiri}
Nojiri S, Odintsov SD.
\newblock Modified $f(R)$ gravity consistent with realistic cosmology: From a matter dominated epoch to a dark energy universe.
\newblock Phys Rev D. 2006 Oct;74:086005.
\newblock Available from: \url{https://link.aps.org/doi/10.1103/PhysRevD.74.086005}.

\bibitem{Amendola_2007b}
Amendola L, Gannouji R, Polarski D, Tsujikawa S.
\newblock Conditions for the cosmological viability of $f(R)$ dark energy models.
\newblock Physical Review D. 2007 Apr;75(8).
\newblock Available from: \url{http://dx.doi.org/10.1103/PhysRevD.75.083504}.

\bibitem{Caprini_2016}
Caprini C, Tamanini N.
\newblock Constraining early and interacting dark energy with gravitational wave standard sirens: the potential of the eLISA mission.
\newblock Journal of Cosmology and Astroparticle Physics. 2016 Oct;2016(10):006–006.
\newblock Available from: \url{http://dx.doi.org/10.1088/1475-7516/2016/10/006}.

\bibitem{oikonomou}
Oikonomou VK, Giannakoudi I.
\newblock A panorama of viable $f(R)$ gravity dark energy models.
\newblock International Journal of Modern Physics D. 2022;31(09):2250075.
\newblock Available from: \url{https://doi.org/10.1142/S0218271822500754}.

\bibitem{ODINTSOV2023137988}
Odintsov SD, Oikonomou VK, Sharov GS.
\newblock Early dark energy with power-law $f(R)$ gravity.
\newblock Physics Letters B. 2023;843:137988.
\newblock Available from: \url{https://www.sciencedirect.com/science/article/pii/S0370269323003222}.

\bibitem{Chatterjee_2024}
Chatterjee A, Roy R, Dey S, Bandyopadhyay A.
\newblock Dynamics of viable $f(R)$ dark energy models in the presence of curvature–matter interactions.
\newblock The European Physical Journal C. 2024 Mar;84(3).
\newblock Available from: \url{http://dx.doi.org/10.1140/epjc/s10052-024-12611-1}.

\bibitem{zhang}
Zhang P.
\newblock Behavior of $f(R)$ gravity in the solar system, galaxies, and clusters.
\newblock Phys Rev D. 2007 Jul;76:024007.
\newblock Available from: \url{https://link.aps.org/doi/10.1103/PhysRevD.76.024007}.

\bibitem{martins}
Martins CF, Salucci P.
\newblock Analysis of rotation curves in the framework of Rn gravity.
\newblock Monthly Notices of the Royal Astronomical Society. 2007 10;381(3):1103-8.
\newblock Available from: \url{https://doi.org/10.1111/j.1365-2966.2007.12273.x}.

\bibitem{stabile}
Stabile A, Capozziello S.
\newblock Galaxy rotation curves in $f(R,\ensuremath{\phi})$ gravity.
\newblock Phys Rev D. 2013 Mar;87:064002.
\newblock Available from: \url{https://link.aps.org/doi/10.1103/PhysRevD.87.064002}.

\bibitem{naik}
Naik AP, Puchwein E, Davis AC, Arnold C.
\newblock Imprints of Chameleon f(R) gravity on Galaxy rotation curves.
\newblock Monthly Notices of the Royal Astronomical Society. 2018 08;480(4):5211-25.
\newblock Available from: \url{https://doi.org/10.1093/mnras/sty2199}.

\bibitem{Starobinsky:1980te}
Starobinsky AA.
\newblock {A New Type of Isotropic Cosmological Models Without Singularity}.
\newblock Phys Lett. 1980;B91:99-102.

\bibitem{joachim}
N\"af J, Jetzer P.
\newblock Gravitational radiation in quadratic $f(R)$ gravity.
\newblock Phys Rev D. 2011 Jul;84:024027.
\newblock Available from: \url{https://link.aps.org/doi/10.1103/PhysRevD.84.024027}.

\bibitem{Zhdanov_2024}
Zhdanov VI, Stashko OS, Shtanov YV.
\newblock Spherically symmetric configurations in the quadratic $f(R)$ gravity.
\newblock Phys Rev D. 2024 Jul;110:024056.
\newblock Available from: \url{https://link.aps.org/doi/10.1103/PhysRevD.110.024056}.

\bibitem{Giamalas}
Gialamas ID, Tamvakis K.
\newblock Bimetric Starobinsky model.
\newblock Phys Rev D. 2023 Nov;108:104023.
\newblock Available from: \url{https://link.aps.org/doi/10.1103/PhysRevD.108.104023}.

\bibitem{Asaka_2016}
Asaka T, Iso S, Kawai H, Kohri K, Noumi T, Terada T.
\newblock Reinterpretation of the Starobinsky model.
\newblock Progress of Theoretical and Experimental Physics. 2016 Dec;2016(12):123E01.
\newblock Available from: \url{http://dx.doi.org/10.1093/ptep/ptw161}.

\bibitem{Gegenberg_2018}
Gegenberg J, Seahra SS.
\newblock Gravitational wave defocussing in quadratic gravity.
\newblock Classical and Quantum Gravity. 2018 jan;35(4):045012.
\newblock Available from: \url{https://dx.doi.org/10.1088/1361-6382/aaa4eb}.

\bibitem{Chakraborty_2021}
Chakraborty S, Gregoris D.
\newblock Cosmological evolution with quadratic gravity and nonideal fluids.
\newblock The European Physical Journal C. 2021 Oct;81(10).
\newblock Available from: \url{http://dx.doi.org/10.1140/epjc/s10052-021-09697-2}.

\bibitem{Nashed_2023}
Nashed GGL, El~Hanafy W.
\newblock Constraining quadratic f(R) gravity from astrophysical observations of the pulsar J0704+6620.
\newblock Journal of Cosmology and Astroparticle Physics. 2023 sep;2023(09):038.
\newblock Available from: \url{https://dx.doi.org/10.1088/1475-7516/2023/09/038}.

\bibitem{Plummer}
Plummer HC.
\newblock On the Problem of Distribution in Globular Star Clusters: (Plate 8.).
\newblock Monthly Notices of the Royal Astronomical Society. 1911 03;71(5):460-70.
\newblock Available from: \url{https://doi.org/10.1093/mnras/71.5.460}.

\bibitem{hernquist}
Hernquist L.
\newblock An Analytical Model for Spherical Galaxies and Bulges.
\newblock apj. 1990 jun;356:359.

\bibitem{Baes_2002}
Baes M, Dejonghe H.
\newblock The Hernquist model revisited: Completely analytical anisotropic dynamical models.
\newblock Astronomy \& Astrophysics. 2002 Sep;393(2):485–497.
\newblock Available from: \url{http://dx.doi.org/10.1051/0004-6361:20021064}.

\bibitem{NFW}
{Navarro} JF, {Frenk} CS, {White} SDM.
\newblock {The Structure of Cold Dark Matter Halos}.
\newblock apj. 1996 May;462:563.

\bibitem{Navarro_1997}
Navarro JF, Frenk CS, White SDM.
\newblock A Universal Density Profile from Hierarchical Clustering.
\newblock The Astrophysical Journal. 1997 Dec;490(2):493–508.
\newblock Available from: \url{http://dx.doi.org/10.1086/304888}.

\bibitem{Jing_2000}
Jing YP.
\newblock The Density Profile of Equilibrium and Nonequilibrium Dark Matter Halos.
\newblock The Astrophysical Journal. 2000 May;535(1):30–36.
\newblock Available from: \url{http://dx.doi.org/10.1086/308809}.

\bibitem{Merritt_2006}
Merritt D, Graham AW, Moore B, Diemand J, Terzić B.
\newblock Empirical Models for Dark Matter Halos. I. Nonparametric Construction of Density Profiles and Comparison with Parametric Models.
\newblock The Astronomical Journal. 2006 Jan;132(6):2685–2700.
\newblock Available from: \url{http://dx.doi.org/10.1086/508988}.

\bibitem{Oman_2015}
Oman KA, Navarro JF, Fattahi A, Frenk CS, Sawala T, White SDM, et~al.
\newblock The unexpected diversity of dwarf galaxy rotation curves.
\newblock Monthly Notices of the Royal Astronomical Society. 2015 Aug;452(4):3650–3665.
\newblock Available from: \url{http://dx.doi.org/10.1093/mnras/stv1504}.

\bibitem{kuzio}
Kuzio~de Naray R, Kaufmann T.
\newblock Recovering cores and cusps in dark matter haloes using mock velocity field observations.
\newblock Monthly Notices of the Royal Astronomical Society. 2011 07;414(4):3617-26.
\newblock Available from: \url{https://doi.org/10.1111/j.1365-2966.2011.18656.x}.

\bibitem{Block}
de~Blok WJG, McGaugh SS, Rubin VC.
\newblock High-Resolution Rotation Curves of Low Surface Brightness Galaxies. II. Mass Models.
\newblock The Astronomical Journal. 2001 nov;122(5):2396.
\newblock Available from: \url{https://dx.doi.org/10.1086/323450}.

\bibitem{Lu}
L\"u H, Perkins A, Pope CN, Stelle KS.
\newblock Spherically symmetric solutions in higher-derivative gravity.
\newblock Phys Rev D. 2015 Dec;92:124019.
\newblock Available from: \url{https://link.aps.org/doi/10.1103/PhysRevD.92.124019}.

\bibitem{planckcollaboration1}
{Planck Collaboration}, {Ade, P  A  R }, {Aghanim, N }, {Armitage-Caplan, C }, {Arnaud, M }, {Ashdown, M }, et~al.
\newblock Planck 2013 results. XXII. Constraints on inflation.
\newblock A\&A. 2014;571:A22.
\newblock Available from: \url{https://doi.org/10.1051/0004-6361/201321569}.

\bibitem{planckcollaboration2}
{Planck Collaboration}, {Ade, P  A  R }, {Aghanim, N }, {Armitage-Caplan, C }, {Arnaud, M }, {Ashdown, M }, et~al.
\newblock Planck 2013 results. XVI. Cosmological parameters.
\newblock A\&A. 2014;571:A16.
\newblock Available from: \url{https://doi.org/10.1051/0004-6361/201321591}.

\bibitem{Wyithe_2001}
Wyithe JSB, Turner EL, Spergel DN.
\newblock Gravitational Lens Statistics for Generalized NFW Profiles: Parameter Degeneracy and Implications for Self‐Interacting Cold Dark Matter.
\newblock The Astrophysical Journal. 2001 Jul;555(1):504–523.
\newblock Available from: \url{http://dx.doi.org/10.1086/321437}.

\bibitem{Lokas_2001}
Lokas EL, Mamon GA.
\newblock Properties of spherical galaxies and clusters with an NFW density profile.
\newblock Monthly Notices of the Royal Astronomical Society. 2001 Feb;321(1):155–166.
\newblock Available from: \url{http://dx.doi.org/10.1046/j.1365-8711.2001.04007.x}.

\bibitem{Kuzio_de_Naray_2009}
Kuzio~de Naray R, McGaugh SS, Mihos JC.
\newblock CONSTRAINING THE NFW POTENTIAL WITH OBSERVATIONS AND MODELING OF LOW SURFACE BRIGHTNESS GALAXY VELOCITY FIELDS.
\newblock The Astrophysical Journal. 2009 Feb;692(2):1321–1332.
\newblock Available from: \url{http://dx.doi.org/10.1088/0004-637X/692/2/1321}.

\bibitem{Dutton_2014}
Dutton AA, Macciò AV.
\newblock Cold dark matter haloes in the Planck era: evolution of structural parameters for Einasto and NFW profiles.
\newblock Monthly Notices of the Royal Astronomical Society. 2014 May;441(4):3359–3374.
\newblock Available from: \url{http://dx.doi.org/10.1093/mnras/stu742}.

\bibitem{Lelli2016SPARC}
Lelli F, McGaugh SS, Schombert JM.
\newblock SPARC: Mass Models for 175 Disk Galaxies with Spitzer Photometry and Accurate Rotation Curves.
\newblock Astronomical Journal. 2016;152:157.

\end{thebibliography}
\end{document}